\RequirePackage{fix-cm}
\documentclass[smallextended]{svjour3}       
\smartqed  
\usepackage{graphicx}
\usepackage{tabularx}

\pdfoutput=1
\usepackage{graphics}
\usepackage{amsmath}
\usepackage{amssymb}
\usepackage{psfrag}
\usepackage{epsfig}
\usepackage{epsf}
\usepackage{float}
\usepackage{slashed}
\allowdisplaybreaks

\newcommand{\fett}[1]{\mbox{\boldmath $#1$}}
\newcommand{\fet}[1]{\mbox{$#1$}}

\newcommand{\beq}{\begin{equation}}
\newcommand{\eeq}{\end{equation}}
\newcommand{\beqa}{\begin{eqnarray}}
\newcommand{\eeqa}{\end{eqnarray}}
\newcommand{\nn}{\nonumber \\ }

\newcommand{\pvec}[1]{\vec{#1}\mkern2mu\vphantom{#1}}

\numberwithin{equation}{section}
\renewcommand\theequation{\arabic{equation}}

\begin{document}

\title{Nuclear electromagnetic currents to fourth order in chiral effective field theory}

\author{H.~Krebs \and
  E.~Epelbaum \and
  U.-G.~Mei{\ss}ner}

\institute{H.~Krebs \at Ruhr-Universit\"at Bochum, Fakult\"at f\"ur Physik und Astronomie,
Institut f\"ur Theoretische Physik II, 
  D-44780 Bochum, Germany \\
\email{hermann.krebs@rub.de}
\and
E.~Epelbaum \at Ruhr-Universit\"at Bochum, Fakult\"at f\"ur Physik und Astronomie,
Institut f\"ur Theoretische Physik II, 
  D-44780 Bochum, Germany \\
  \email{evgeny.epelbaum@rub.de}
\and  
U.-G.~Mei{\ss}ner \at 
Helmholtz-Institut f\"ur Strahlen- und Kernphysik and Bethe Center
for Theoretical Physics, Universit\"at Bonn, D-53115 Bonn, Germany;
Institute for Advanced Simulation, Institut f\"ur Kernphysik, and
J\"ulich Center for Hadron Physics, Forschungszentrum J\"ulich,
D-52425 J\"ulich, Germany; 
Tbilisi State  University,  0186 Tbilisi, Georgia \\
\email{meissner@hiskp.uni-bonn.de}
}

\date{Received: date / Accepted: date}

\maketitle
\begin{abstract}
Recently, we have shown that the continuity equation for the nuclear 
vector and axial current operators acquires additional terms 
if the latter depend on the energy transfer. We analyze
in detail the electromagnetic single-nucleon four-current operators
and verify the validity of the modified continuity equation
for all one- and two-nucleon contributions up to fourth order in the chiral expansion.
We also derive, for the first time, the leading
contribution to the three-nucleon charge operator which appears at
this order. Our study  completes the derivation of the electroweak
nuclear currents to fourth order in the chiral expansion.  
\end{abstract}

\section{Introduction}
\def\theequation{\arabic{section}.\arabic{equation}}
\label{sec:intro}

Chiral effective field theory has been extensively used in the past
decades to derive nuclear forces and the corresponding vector and
axial charge and current operators. Presently, the two-nucleon force
contributions have been worked out completely up through fifth order
$Q^5$ of the chiral expansion, i.e.~up to N$^4$LO
\cite{Entem:2014msa,Epelbaum:2014sza,Entem:2015xwa,Reinert:2017usi,Entem:2017gor}. Here and in
what follows, the expansion parameter is defined as $Q \in \{ M_\pi /
\Lambda_b, \; p/\Lambda_b \}$, where $p$ refers to four-
(three-)momenta of external pions (nucleons), while $\Lambda_b$ is the
breakdown scale of the chiral expansion. In the Goldstone boson and
single-baryon sectors one usually assumes $\Lambda_b = \Lambda_\chi \sim
M_\rho \sim 4 \pi F_\pi \sim 1$~GeV \cite{Manohar:1983md}. In the
few-nucleon sector, the breakdown scale is
estimated to be of the order of $\Lambda_b \sim 600$~MeV \cite{Epelbaum:2014efa,Furnstahl:2015rha}.    
The expressions for the three- and four-nucleon forces 
up to fourth order, i.e.~$Q^4$ or N$^3$LO, calculated using
dimensional regularization can be found in
Refs.~\cite{vanKolck:1994yi,Epelbaum:2004fk,Epelbaum:2006eu,Ishikawa:2007zz,Bernard:2007sp,Bernard:2011zr}.
Most of the N$^4$LO
contributions to the three-nucleon force have also been worked out
\cite{Krebs:2012yv,Krebs:2013kha,Epelbaum:2014sea,Girlanda:2011fh}. 

Nuclear electroweak current operators have also attracted considerable
attention. Starting from the pioneering (but incomplete and
kinematically restricted) studies by Park et al.~\cite{Park:1993jf,Park:1995pn}, the exchange
electromagnetic currents have been worked out to the leading
one-loop order $Q$, i.e.~N$^3$LO\footnote{The dominant contributions
  to the single- and two-nucleon operators appear at orders
  $Q^{-3}$ and $Q^{-1}$, respectively. When the nucleon mass $m$ is
  counted according to $m \sim \Lambda_b^2/Q$~\cite{Weinberg:1991um} as done here and
  commonly employed in few-nucleon studies, see
  e.g.~Refs.~\cite{Epelbaum:2014sza,Reinert:2017usi,Epelbaum:2014efa,Bernard:2007sp,Bernard:2011zr,Krebs:2012yv,Krebs:2013kha,Epelbaum:2014sea},
  no corrections to the charge and current operators appear at
  order $Q^{-2}$. Thus, the terms LO, NLO, N$^2$LO and N$^3$LO refer
  to the contributions at orders $Q^{-3}$, $Q^{-1}$, $Q^{0}$ and
  $Q$, respectively. Notice that the authors of
  Refs.~\cite{Pastore:2008ui,Pastore:2009is,Pastore:2011ip} employ the
  counting scheme with $m \sim \Lambda_b$ leading to a different
  expansion pattern. },
using the method of unitary
transformation (UT) \cite{Epelbaum:1998ka,Epelbaum:2007us,Epelbaum:2008ga} in Refs.~\cite{Kolling:2009iq,Kolling:2011mt} and time-ordered
perturbation theory in
Refs.~\cite{Pastore:2008ui,Pastore:2009is,Pastore:2011ip}, see also
Ref.~\cite{Hoferichter:2015ipa} for a related discussion.
In fact, the UT method was first used in this context in Ref.~\cite{Walzl:2001vb}.
The main idea behind the method of
UT is to find a unitary operator in the
pion-nucleon Fock space which decouples its purely nucleonic subspace,
called $\eta$-space,
from the rest of the Fock space. Nuclear forces are then identified
with the resulting decoupled Hamiltonian which acts on the
nucleonic subspace. Notice that it is
always possible to perform additional UTs on the
$\eta$-space which affect the off-shell behavior of the nuclear
potentials but do not change observable quantities. 
This unitary ambiguity was found to be considerably reduced if one
demands renormalizability of the nuclear potentials \cite{Epelbaum:2006eu,Epelbaum:2007us}. 
While the on-shell scattering amplitude calculated in chiral effective field
theory is defined unambiguously and guaranteed to be renormalizable,
this does not hold 
anymore for irreducible (i.e.~non-iterative) contributions alone,
which are associated with nuclear forces and current operators.
Renormalizability of nuclear potentials can,
however, be enforced by performing suitable UTs on the $\eta$-space.  
For a broad class of UTs considered in Refs.~\cite{Epelbaum:2006eu,Epelbaum:2007us,Bernard:2007sp},
the resulting renormalized static contributions to the nuclear forces up to
N$^4$LO were found to be determined unambiguously, while the
leading 
relativistic corrections depend on two arbitrary phases $\bar
\beta_8$ and $\bar \beta_9$
\cite{Bernard:2011zr,Epelbaum:2014efa}, see also
\cite{Friar:1999sj} and references therein for a related earlier discussion
of this kind of unitary ambiguity.

In the presence of external
classical sources, the UTs  determined in the strong
sector as outlined above allow one to derive the corresponding charge
and current operators. To enforce
renormalizability also for the four-current operators one, however, needs to
employ a more general class of $\eta$-space UTs, whose generators
involve a single insertion of an external classical sources. At the
order considered they are  parametrized via
\cite{Kolling:2011mt}\footnote{The minus sign is missing in
  Eq.~(4.3) of Ref.~\cite{Kolling:2011mt}
  and  for the quantities $\delta J_{c2, \, c5, \, c6, \, c7}$ in Eq.~(4.6) of that paper.}
\beq
\label{betas}
S = - \sum_{i=1}^7  \bar\beta_i S_i \,,
\eeq
where $\bar\beta_i \in \mathbb{R}$ are the transformation ``angles''
while the corresponding anti-hermitian generators $S_i$  have the form 
\beqa
S_1 &=& \eta \left[ J_{02}^{(-1)} \frac{\lambda^2}{E_\pi^2} H_{22}^{(2)} \; - \;
  H_{22}^{(2)} \frac{\lambda^2}{E_\pi^2} J_{02}^{(-1)} \right] \eta\,, \nn[3pt]
S_2 &=& \eta \left[ H_{21}^{(1)} \frac{\lambda^1}{E_\pi^2} J_{20}^{(-1)}
  \frac{\lambda^1}{E_\pi} H_{21}^{(1)}\; - \;
  H_{21}^{(1)} \frac{\lambda^1}{E_\pi} J_{20}^{(-1)} 
\frac{\lambda^1}{E_\pi^2} H_{21}^{(1)} \right]\eta\,, \nn[3pt]
S_3 &=& \eta \left[ J_{20}^{(-1)} \eta H_{21}^{(1)}
  \frac{\lambda^1}{E_\pi^3} H_{21}^{(1)}\; - \;
  H_{21}^{(1)} \frac{\lambda^1}{E_\pi^3} H_{21}^{(1)}\eta 
 J_{02}^{(-1)} \right]\eta\,, \nn[3pt]
S_4 &=& \eta\left[ J_{02}^{(-1)} \frac{\lambda^2}{E_\pi^2}H_{21}^{(1)}
    \frac{\lambda^1}{E_\pi}H_{21}^{(1)} - H_{21}^{(1)}
    \frac{\lambda^1}{E_\pi}H_{21}^{(1)} \frac{\lambda^2}{E_\pi^2}J_{02}^{(-1)}
  \right]\eta  \,, \nn[3pt]
S_5 &=& \eta\left[ J_{02}^{(-1)} \frac{\lambda^2}{E_\pi}H_{21}^{(1)}
    \frac{\lambda^1}{E_\pi^2}H_{21}^{(1)} - H_{21}^{(1)}
    \frac{\lambda^1}{E_\pi^2}H_{21}^{(1)} \frac{\lambda^2}{E_\pi}J_{02}^{(-1)}
  \right]\eta  \,, \nn[3pt]
S_6 &=& \eta\left[ H_{21}^{(1)} \frac{\lambda^1}{E_\pi}J_{02}^{(-1)} 
    \frac{\lambda^1}{E_\pi^2}H_{21}^{(1)} - H_{21}^{(1)}
    \frac{\lambda^1}{E_\pi^2}J_{02}^{(-1)} \frac{\lambda^2}{E_\pi}H_{21}^{(1)}
  \right]\eta  \,, \nn[3pt]
S_7 &=& \eta\left[ H_{21}^{(1)} \frac{\lambda^1}{E_\pi^2}J_{21}^{(0)} - J_{21}^{(0)}
    \frac{\lambda^1}{E_\pi^2}H_{21}^{(1)}
  \right]\eta  \,.
\label{unitS}
\eeqa
Here, $\eta$ and $\lambda^i$ denote the projection operators onto the
purely nucleonic part of the Fock space and Fock-space components with
$i$ pions, respectively. Further, $E_{\pi} =\sum_i \sqrt{\vec p_i
  ^{\, 2} + M_\pi^2}$ is the total energy of pions with three-momenta
$\vec p_i$ in the
intermediate $\lambda^i$-state while  $H_{ab}^{(\kappa )}$
($J_{ab}^{(\kappa )}$) refer to the vertices in the effective chiral
Hamiltonian with $a$ nucleon and $b$ pion fields without (with a
single insertion of) external electromagnetic sources.  The
superscript $\kappa$ is related to the canonical field dimension of
the fields via $\kappa = d + (3/2) n + p - 4$, where $d$, $n$ and $p$
denote the number of derivatives or $M_\pi$-insertions and the
nucleon and pion field operators, respectively. In the method of UT,
all contributions to the nuclear forces and currents are written as
time-ordered sequences of vertices and energy denominators.
A contribution involving $N$ vertices of type
$\kappa_i$, $i=1, \ldots N$, appears at order $Q^\nu$ with $\nu = \sum_i \kappa_i
-2$ ($\nu = \sum_i \kappa_i
-3$) for the forces (currents).
More details on the notation and the method of UT can be found in
Ref.~\cite{Epelbaum:2006eu,Epelbaum:2007us}, while the
explicit form of the relevant terms in the effective Lagrangian and
Hamiltonian is given in Ref.~\cite{Kolling:2011mt}. For the sake of
completeness, we also give the expressions for the generators of the 
already mentioned $\eta$-space UTs which parametrize the unitary ambiguity of the
leading relativistic corrections to the nuclear forces and currents:  
\beq
  S^\prime =  \bar\beta_8 S_8 + \bar\beta_9 S_9\,, 
\eeq
with the operators $S_{8,9}$ given by   
\begin{eqnarray}
  S_8 & = & \eta \left[ \tilde H_{20}^{(2)} \eta H_{21}^{(1)}\frac{\lambda_1}{E_\pi^3}
  H_{21}^{(1)} - H_{21}\frac{\lambda_1}{E_\pi^3}
  H_{21}^{(1)} \eta \tilde H_{20}^{(2)} \right]\eta\,, \nn
S_9 & = & \eta \left[ \tilde H_{21}^{(3)} \frac{\lambda^1}{E_\pi^2}H_{21}^{(1)} -
  H_{21}^{(1)} \frac{\lambda^1}{E_\pi^2} \tilde H_{21}^{(3)}\right] \eta\,.
\end{eqnarray}
Here, $\tilde H_{ab}^{(\kappa )}$ denote the $1/m$-corrections to
the corresponding vertices.   

The UTs listed in Eqs.~(\ref{betas},\ref{unitS}) induce contributions to the
electromagnetic charge and current operators at N$^3$LO and higher
chiral orders which restore their renormalizability upon a suitable
choice of $\bar \beta_i$. More precisely, it was found in
Ref.~\cite{Kolling:2011mt} that the expressions for the two-nucleon
charge and current operators do not depend on the parameters $\bar
\beta_2$,  $\bar \beta_7$ and $\bar \beta_5 -\bar \beta_6$. Further,
renormalizability of the one-loop contributions to the one-pion
exchange current operator, calculated  in dimensional regularization,
was found to require the choice
\beq
\bar \beta_1 = 1, \quad \quad 
\bar \beta_4 - 3 \bar \beta_5 - 3 \bar \beta_6 = 2\,, 
\eeq
while the parameter $\bar \beta_3$ was set to $\bar \beta_3=0$.
With this choice, the one-loop contributions to the electromagnetic
charge and current operators still depend on an arbitrary phase $\bar
\beta_4$, for which the value of $\bar \beta_4 = -1$ was adopted  in
Ref.~\cite{Kolling:2011mt}.\footnote{Requiring, in addition to 
  renormalizability of the single- and two-nucleon current operators,
  also factorizability of
  the exchanged pions in the three-nucleon charge operator, see
  \cite{Krebs:2016rqz} for more details, leads to the stronger
  constraints on the phases, namely $\bar \beta_1 = 1$, $\bar \beta_4
  = -1$ and $\bar \beta_5 = \bar \beta_6 = -1/2$.} 

In Ref.~\cite{Krebs:2016rqz}, the framework outlined above was extended to derive
the nuclear axial charge and current operators. Similarly to the vector
currents, renormalized expressions could only be obtained by
taking into account additional $\eta$-space UTs whose generators
depend on the external axial sources and have been parametrized in
terms of the real phases $\alpha_i^{ax}$, $i=1, 2, \ldots,
33$. While renormalizability alone does not completely eliminate the
resulting unitary ambiguity, the requirement of the pion-pole
contributions to the axial current to match the corresponding terms in
the nuclear potentials was shown to result in unique expressions.
Ref.~\cite{Krebs:2016rqz} also provides a detailed discussion of
the subtleties associated with the explicit time dependence of the
additional UTs due to the dependence of their generators on the
external classical sources. It is shown in that paper that the explicit time dependence of
the UTs is, in general, expected to yield contributions which
depend on the energy transfer $k_0$ of the external source. Also the
continuity equations, which are manifestations of
gauge invariance and the
chiral symmetry, get modified and take a more general form that
involves the energy-dependent contributions to the charge and current
operators, see Eq.~(2.42) of Ref.~\cite{Krebs:2016rqz}. The validity
of the continuity equation for the axial current was explicitly
verified in that paper. For a related recent work on the nuclear axial
currents in the framework of time-ordered perturbation theory see
Refs.~\cite{Baroni:2015uza}. 

The purpose of this study is to complete the derivation of the electromagnetic
currents at N$^3$LO initiated in \cite{Kolling:2009iq,Kolling:2011mt}
and to investigate the implications of the
findings of Ref.~\cite{Krebs:2016rqz}. In particular,  we discuss
in detail the expressions for the single-nucleon 
charge and current operators including the relevant relativistic
corrections and provide their parametrization in
terms of the corresponding nucleon form factors. We also work out the
three-nucleon contributions to the charge operator, which completes the derivation of the
nuclear electromagnetic currents at N$^3$LO. Last but not least, we explicitly
verify the validity of the continuity equation for all considered 
classes of diagrams. Notice that since  single-nucleon contributions
were not considered in our earlier papers
\cite{Kolling:2009iq,Kolling:2011mt}, the continuity equation could
so far only be 
verified explicitly for the static two-pion exchange terms. 

Our paper is organized as follows. In Sec.~\ref{sec:1N} we discuss the single-nucleon
charge and current operators. Special attention is paid to the
energy-transfer-dependent contributions, which lead to the already mentioned
modified form of the continuity equation. In  Sec.~\ref{sec:continuity:OPE} we use the
derived single-nucleon current to explicitly verify the validity of the 
continuity equation for the one-pion exchange (OPE) and 
short-range two-nucleon operatiors.
Next, in Sec.~\ref{sec:3N} we derive the leading
three-nucleon contributions to the electromagnetic charge operator. We observe that
there are no three-nucleon current contributions up to the considered
order. The main results of our paper are summarized in Sec.~\ref{sec:summary}.

\section{Single nucleon electromagnetic current}
\def\theequation{\arabic{section}.\arabic{equation}}
\label{sec:1N}

Here and in what follows, we stay as close as possible to the
notation employed in Ref.~\cite{Krebs:2016rqz}. In particular, the one-, two- and
three-nucleon four-current operators 
are defined according to
\beqa
\langle \pvec p ' | \hat{\fet V}^{\mu }_{\rm 1N} | \vec
p \, \rangle &=:& \delta^{(3)} ( \pvec p '  - \vec p
- \vec k )    \fet V^{\mu }_{\rm 1N} \,, \nn
\langle \pvec p_1 ' \pvec p_2 ' | \hat{\fet V}^{\mu }_{\rm 2N} | \vec
p_1 \vec p_2 \rangle &=:& (2\pi)^{-3}\delta^{(3)} ( \pvec p_1  ' + \pvec p_2 ' - \vec p_1
- \vec p_2 - \vec k )    \fet V^{\mu }_{\rm 2N} \,, \\
  \langle \pvec p_1 ' \pvec p_2 ' \pvec p_3 ' | \hat{\fet V}^{\mu}_{\rm 3N} | \vec
p_1 \vec p_2 \vec p_3\rangle &=:& (2\pi)^{-6}\delta^{(3)} ( \pvec p_1  ' + \pvec p_2 ' + \pvec p_3 '- \vec p_1
- \vec p_2 - \vec p_3 - \vec k ) \fet  V^{\mu}_{\rm 3N} \,,
\nonumber
\eeqa
where $\vec p_i$ ($\pvec p_i '$) denotes the incoming (outgoing)
momentum of nucleon $i$ while $\vec k$ is the momentum of the external
electromagnetic 
source. The electromagnetic
current operators we are interested in here are linear combinations of
isoscalar and isovector quantities. For the sake of compactness, we
refrain from explicitly showing isospin indices except for the isospin
Pauli matrices $\fett \tau$. Further, $\hat X$ means that the quantity
$X$ is to be regarded as an 
 operator rather than matrix element with respect to the nucleon
 momenta. 
Finally, we employ throughout our work the same choice of the unitary
 phases $\bar \beta_i$ as adopted in Ref.~\cite{Kolling:2011mt} and
 explained in the previous section.

\subsection{Parametrization of the single nucleon current in terms of
  the form factors}
\label{sec:1N:ffparametrization}

Single-nucleon current operators can be conveniently parametrized in
terms of the Dirac and Pauli electromagnetic form
factors of the nucleon $F_1 (Q^2)$ and $F_2 (Q^2)$, respectively.
In the relativistic kinematics, the on-shell single-nucleon
current is given by
\beqa
\fet{\mathcal{V}}_{{\rm 1N}}^\mu&=& \frac{e}{2 m} \,
\bar{u}(p^\prime\,)\bigg[\gamma^\mu {\fet F}_1(Q^2) +
\frac{i}{2 m}\sigma^{\mu\nu} k_\nu {\fet F}_2(Q^2)\bigg]u(p),
\eeqa
where the $\gamma^\mu$ are Dirac matrices, 
$\sigma^{\mu\nu}=i[\gamma^\mu,\gamma^\nu]/2$,
$k=p^\prime-p$ is the  four-momentum transfer while $Q^2=-k_\mu
k^\mu$. The Dirac spinors are normalized according to 
\beqa
\bar{u}(p) u(p) &=&2 m.
\eeqa
As already pointed out above, we suppress the isospin indices to
simplify the notation.
In the following, we will express all our results in terms of the 
Sachs electric and magnetic form factors
\beqa
{\fet G}_E(Q^2)&=&{\fet F}_1(Q^2) - \frac{Q^2}{4 m^2} {\fet F}_2(Q^2),
\nonumber \\
{\fet G}_M(Q^2)&=&{\fet F}_1(Q^2) + {\fet F}_2(Q^2).
\eeqa
The $1/m$-expansion of the non-relativistically normalized electromagnetic
current is defined according to 
\beq
\sqrt{\frac{m}{E_{p^\prime}}} \fet{\mathcal{V}}_{{\rm 1N}}^\mu \sqrt{\frac{m}{E_{p}}}=
\chi^{\prime\dagger}\bigg({\fet V}_{{\rm 1N:\, static}}^\mu +  {\fet
  V}_{{\rm 1N:} \, 1/m}^\mu + {\fet V}_{{\rm 1N:} \,  1/m^2}^\mu +
{\cal O}\bigg(\frac{1}{m^3}\bigg)\bigg)\chi,
\eeq
where $E(p)=\sqrt{m^2 + \vec{p}^{\,2}}$, while $\chi^\prime$ and $\chi$
refer to Pauli spinors.
In terms of the Sachs form factors,  the charge operator is
parametrized via
\beqa
{\fet V}_{{\rm 1N:\, static}}^0&=&e {\fet G}_E(Q^2),\nn
{\fet V}_{{\rm 1N:} \, 1/m}^0&=&
\frac{i\,e}{2m^2}\vec{k}\cdot(\vec{k}_1\times\vec{\sigma}) {\fet G}_M(Q^2),\nn
{\fet V}_{{\rm 1N:}\, 1/m^2}^0&=& -\frac{e}{8
  m^2}\big[Q^2+2\,i\,\vec{k}\cdot(\vec{k}_1\times\vec{\sigma}) \big]
{\fet G}_E(Q^2),\label{V1N:Charge:nonrel}
\eeqa
while the current operator  is given by
\beqa
\vec{{\fet V}}_{{\rm 1N:\, static}}&=&- \frac{i\,e}{2
  m}\vec{k}\times\vec{\sigma}\, {\fet G}_M(Q^2),\nn
\vec{{\fet V}}_{{\rm 1N:} \, 1/m}&=&\frac{e}{m}\vec{k}_1\,{\fet G}_E(Q^2),\label{V1N:Current:nonrel}\nn
\vec{{\fet V}}_{{\rm 1N:} \, 1/m^2}&=&\frac{e}{16
  m^3}\bigg[i\,\vec{k}\times\vec{\sigma}(2\vec{k}_1^{\,2}+Q^2)+2\,i\,\vec{k}\times\vec{k}_1\,\vec{k}_1\cdot\vec{\sigma}
\\
&+& 2
\vec{k}_1(i\,\vec{k}\cdot(\vec{k}_1\times\vec{\sigma})+Q^2)
-2\,\vec{k}\,\vec{k}\cdot\vec{k}_1
+ 6\,i\,\vec{k}_1\times\vec{\sigma}\,\vec{k}\cdot\vec{k}_1\bigg]{\fet
  G}_M(Q^2),
\nonumber
\eeqa
where
$\vec{k}_1=(\vec{p}^{\,\prime}+\vec{p} \, )/2$ and $\vec \sigma$ refers
to the Pauli matrices in spin space. 
The powers of the  nucleon mass in
Eqs.~(\ref{V1N:Charge:nonrel},\ref{V1N:Current:nonrel})
can be traced back to the convention, according to which the nonrelativistic expansion of the
electric and magnetic Sachs form factors starts from terms of orders
$m^0$ and $m^1$, respectively.  Notice further that replacing $Q^2$ by
$\vec{k}^2$  does not affect
Eqs.~(\ref{V1N:Charge:nonrel},\ref{V1N:Current:nonrel}) since
$Q^2=\vec{k}^2+{\cal O}(1/m^2)$.

Before discussing the results for the charge and current operators
in the method of UT, which may generally be expected to differ
from the on-shell expressions given in Eqs.~(\ref{V1N:Charge:nonrel}) and (\ref{V1N:Current:nonrel})
by off-shell terms, it is instructive to address the validity of the
continuity equation for the on-shell contributions alone. The general form
of the continuity equation for the vector current at the considered
order in the chiral expansion can be written as \cite{Krebs:2016rqz}
\beq
\vec{k}\cdot{\hat{\vec{\fet V}}}(\vec{k},0)-\bigg[\hat{{\fet{H}}},\, \hat{\fet
  V}_0(\vec{k},0)-\frac{\partial}{\partial
  k_0}\bigg(\vec{k}\cdot{\hat{\vec{\fet
      V}}}(\vec{k},k_0)-\big[\hat{{\fet{H}}},\, \hat{\fet
  V}_0(\vec{k},k_0)\big]\bigg)\bigg]= 0,\label{Continuityeq:general}
\eeq 
where ${\fet H}$ denotes the strong part of the nuclear Hamiltonian.
The longitudinal part of
the on-shell single nucleon current operator is given by
\beqa
\vec{k}\cdot{\vec{\fet V}}_{{\rm 1N:\, static}}&=&0,\nn
\vec{k}\cdot{\vec{\fet V}}_{{\rm
    1N:}\, 1/m}&=&\frac{\vec{k}\cdot\vec{k}_1}{m} e\, {\fet G}_E(Q^2),
\nn
\vec{k}\cdot{\vec{\fet V}}_{{\rm
    1N:} \, 1/m^2}&=&i\,\frac{\vec{k}\cdot\vec{k}_1}{2
  m^3}\vec{k}\cdot(\vec{k}_1\times\vec{\sigma}) e\, {\fet G}_M(Q^2),
\eeqa
while the commutator of the kinetic energy with the on-shell charge
operator reads
\beqa
\bigg[\frac{\hat{\vec{p}}^{\, 2}}{2 m}, \, \hat{\fet V}_{{\rm
    1N: \, static}}^0\bigg]&=& \frac{\vec{k}\cdot\vec{k}_1}{m} e\, {\fet
  G}_E(Q^2), \nn
\bigg[\frac{\hat{\vec{p}}^{\, 2}}{2 m}, \, \hat{\fet V}_{{\rm
    1N:}\, 1/m}^0\bigg]&=& i\,\frac{\vec{k}\cdot\vec{k}_1}{2
  m^3}\vec{k}\cdot(\vec{k}_1\times\vec{\sigma}) e\, {\fet G}_M(Q^2).
\eeqa
Thus, if there is no dependence on the energy transfer, the parametrized on-shell
current satisfies the continuity equation in the single-nucleon
sector. This is, however, not the only possibility. If, for example,
the  current operator has an off-shell longitudinal component given by
\beqa
\vec{\fet V}_{{\rm
    1N: \, off-shell}}&=&\vec{k}\bigg(k_0-\frac{\vec{k}\cdot\vec{k}_1}{m}\bigg)
\frac{e}{Q^2}\, \fet X\label{singleN:offshell0}\,,
\eeqa
where $\fet X$ is an arbitrary function of momenta, the single-nucleon electromagnetic current
\beqa
{\fet V}_{{\rm 1N}}^0 &=& {\fet V}_{{\rm 1N:\, static}}^0 + {\fet
  V}_{{\rm 1N:}\, 1/m}^0 + {\fet V}_{{\rm 1N:}\, 1/m^2}^0,\nn
\vec{{\fet V}}_{{\rm 1N}} &=& \vec{{\fet V}}_{{\rm 1N:\, static}} + \vec{{\fet
  V}}_{{\rm 1N:}\, 1/m} + \vec{{\fet V}}_{{\rm 1N:}\, 1/m^2} +\vec{\fet V}_{{\rm
    1N:\, off-shell}}\label{singleN:general}
\eeqa
still satisfies the continuity equation~(\ref{Continuityeq:general})
since 
\beqa
\vec{k}\cdot{\vec{\fet V}}_{{\rm
    1N:\, off-shell}}(\vec{k},0)&=&-\frac{\vec{k}\cdot\vec{k}_1}{m} e \, \fet X
,
\eeqa
and
\beqa
\bigg[\frac{\hat{\vec{p}}^{\, 2}}{2 m}, \, \frac{\partial}{\partial k_0}\vec{k}\cdot\hat{\vec{{\fet V}}}_{{\rm
    1N: \, off-shell}}(\vec{k},k_0)\bigg]&=&\frac{\vec{k}\cdot\vec{k}_1}{m}
e
\, \fet X\,.
\eeqa

We now turn to the method of UT. Performing explicit calculations, we
find that the on-shell expressions for the nuclear charge and current
operators to have the same form
as given in Eqs.~(\ref{V1N:Charge:nonrel}) and
(\ref{V1N:Current:nonrel}). Using the same choice of the additional
$\eta$-space UTs as adopted in Ref.~\cite{Kolling:2011mt} and
described in
Sec.~\ref{sec:intro}, we, however, find an additional off-shell
contribution to the single-nucleon current operator which depends on
the phase $\propto \bar
\beta_6$, see Eq.~(\ref{unitS}). 
Moreover, using 
dimensional regularization to calculate the loop integrals at order
$Q$, we obtain a
divergent contribution to the current of the form  
\beqa
\vec{{\fet V}}_{{\rm 1N: \, singular}}^{(Q)}&=& - \vec{k}
\,\bigg(k_0-\frac{\vec{k}\cdot\vec{k}_1}{m}\bigg) \tau_3\frac{2\bar{\beta}_6+1}{d-4}\frac{5\,e\,g_A^2}{6 (4\pi
    F_\pi)^2}\,,
  \eeqa
where $d$ denotes the number of space-time dimensions.
Thus, renormalizability of the current operators requires taking
$\bar{\beta}_6=-1/2$, and  the
finite pieces of
single-nucleon current operator does explicitly depend 
on the energy transfer $k_0$. The corresponding off-shell contribution
to the current operator can be expressed in terms of the form factors as
\beqa
\label{singleN:offshell}
\vec{\fet V}_{{\rm
    1N:\, off-shell}}&=&\vec{k}\bigg(k_0-\frac{\vec{k}\cdot\vec{k}_1}{m}\bigg)
\frac{e}{Q^2}\\
&\times & \bigg[\big({\fet G}_E(Q^2)-{\fet
  G}_E(0)\big)+\frac{i}{2m^2}\vec{k}\cdot(\vec{k}_1\times\vec{\sigma})\big({\fet
  G}_M(Q^2)-{\fet G}_M(0)\big)\bigg].
\nonumber
\eeqa
Clearly, the unitary transformation $\propto \bar \beta_6$ in
Eq.~(\ref{unitS}) only induces terms in $\vec{\fet V}_{{\rm
    1N: \, off-shell}} $ up to order $Q$.
In Appendix~\ref{unitarytr:singenucleon:general} we, however, provide an
explicit form of the unitary
transformation in the Hilbert space, which exactly 
generates the off-shell terms given in Eq.~(\ref{singleN:offshell}).

\subsection{Chiral expansion of the  Sachs form factors}
\label{sec:1N:OrderbyOrderff}

In order to verify the validity of the continuity equation 
 we provide in this section explicit
expressions for the single-nucleon form factors at lowest orders
in the chiral expansion, as already employed in Ref.~\cite{Walzl:2001vb}.
We refer the reader to
Ref.~\cite{Kubis:2000zd,Schindler:2005ke} and references therein for a detailed analysis of the
electromagnetic form factors of the nucleon in relativistic
formulations of baryon chiral perturbation theory.  We
employ in this section the standard power counting used in 
the single-nucleon heavy baryon approach, where the nucleon mass is
treated on the same footing as the chiral symmetry breaking scale
$m\sim\Lambda_\chi$. The corresponding powers of the soft scale
relative to the dominant contribution will be written in square 
brackets to avoid a possible confusion with the counting scheme
employed throughout the rest of this paper 
as explained in Sec.~\ref{sec:intro}. For what concerns the
relativistic corrections, we only list below the contributions which
are relevant to the derivation of the current operators at the
considered order.
For the sake of simplicity, we do note differentiate between
the various quantities in the chiral limit and their physical values here.
\begin{itemize}
\item
The leading contributions to the form factors ${\fet G}_E(Q^2)$ and ${\fet G}_M(Q^2)$
have the form
\beqa
{\fet G}_E^{[Q^{0}]}(Q^2)&=&\frac{1}{2}\big(1+\tau_3\big),\nn
{\fet G}_M^{[Q^{0}]}(Q^2)&=&\frac{1}{2}\big(1+\kappa_s +
\big(1+\kappa_v\big)\tau_3\big)\,,
\eeqa
where $\kappa_s$ and $\kappa_v$ refer to the isoscalar and isovector anomalous magnetic
moments of the nucleon, respectively. 
\item
At order $Q^{1}$, there are only loop contributions to the magnetic form factor
\beqa
{\fet G}_E^{[Q]}(Q^2)&=&0, \nn {\fet
  G}_M^{[Q]}(Q^2)&=&\frac{m g_A^2 \tau_3}{16 F_\pi^2
  \pi}\big[M_\pi-(4M_\pi^2+Q^2)A(| \vec k \,|)\big]\,,
\eeqa
where $g_A$, $F_\pi$ and $M_\pi$ refer to the nucleon axial-vector
coupling, pion decay constant and pion mass, respectively. Further,
the loop function $A (| \vec k \,|)$ is given by
\beq
A (k) = \frac{1}{2 | \vec k \,|} \arctan \frac{| \vec k \,|}{2 M_\pi}\,.
\eeq
\item
At order $Q^{2}$ one has further contributions to both electromagnetic
form factors. These read
\beqa
{\fet G}_E^{[Q^{2}]}(Q^2)&=&\frac{1}{6 (4\pi F_\pi)^2}\big[4(1+2
g_A^2)M_\pi^2 + (1+5 g_A^2)Q^2\big]L(| \vec k \,|) \tau_3 \nn
&+&\frac{\tau_3}{36 (4\pi F_\pi)^2}\big[-24 (1+2 g_A^2) M_\pi^2 -
Q^2(5 + 13 g_A^2)\big] \nn
&+& Q^2 (2 \bar{d}_7 + \bar{d}_6
\tau_3),\nn
{\fet G}_M^{[Q^{2}]}(Q^2)&=& m\bigg[\frac{c_4\tau_3}{9(4\pi
  F_\pi)^2}\big(24 M_\pi^2 (L(| \vec k \,|)-1)+Q^2(6 L(| \vec k \,|)-5)\big)\nn
&-& 2Q^2(2
\bar{e}_{54} + \bar{e}_{74}\tau_3)\bigg]
+\frac{\tau_3}{36(4 \pi
  F_\pi)^2}\big[24(4 g_A^2 - 1) M_\pi^2 (1-L(| \vec k \,|)) \nn
&-& Q^2(5-23 g_A^2 +
6(7 g_A^2 - 1) L(| \vec k \,|))\big]\,,
\eeqa
with the loop function $L(| \vec k \,|)$ given by 
\beq
L(| \vec k \,|) = \frac{\sqrt{\vec k^2 + 4 M_\pi^2}}{| \vec k \,|}
\log \frac{\sqrt{\vec k^2 + 4
    M_\pi^2} + | \vec k \,|}{2 M_\pi}\,.
\eeq
Here, $c_i$, $\bar d_i$ and $\bar e_i$ are the low-energy constants
(LECs) from the second-, third- and fourth-order pion-nucleon effective
chiral Lagrangian, see Ref.~\cite{Fettes:2000gb} for details.  
\item
At order $Q^3$ one has to take into account the leading two-loop
contributions, whose calculation goes beyond the scope of our study,
as well as the additional relativistic corrections at the one-loop
level given by
\beqa
{\fet G}_E^{[Q^3]}(Q^2)&=&\frac{g_A^2 \tau_3}{32 \pi F_\pi^2
  m}\big[M_\pi(M_\pi^2-2Q^2)-(2M_\pi^2+Q^2)^2 A(| \vec k \,|)\big],\nn
{\fet G}_M^{[Q^3]}(Q^2)&=&\frac{g_A^2 Q^2}{512 \pi F_\pi^2 m}\big[2 (16
M_\pi^2 + 5 Q^2)A(| \vec k \,|)\tau_3 \nn
&+& (3(1+\kappa_s) +(11 - \kappa_v)\tau_3) M_\pi\big].
\eeqa
\end{itemize}
Notice that the chiral expansion of the electromagnetic form
factors of the nucleon is known to converge rather slowly. The form
factors are known to be largely driven by vector mesons \cite{Kubis:2000zd,Schindler:2005ke},
which are not included as explicit degrees of freedom in the approach we are
using. In this sense, it is certainly more advantageous from the
phenomenological point of view to employ
empirical parametrizations of the single-nucleon form factors in 
the single-nucleon charge and current operators
when performing  few-nucleon calculations, rather then to 
strictly rely on their chiral expansion. Such an approach is, in fact,
adopted in most of the studies in the few-nucleon sector, see e.g. the
early work in Ref.~\cite{Phillips:1999am}.

\section{Current conservation}
\label{sec:continuity:OPE}

Having derived the single-nucleon contributions we are now in the
position to explicitly demonstrate current conservation by verifying
the validity of the continuity equation (\ref{Continuityeq:general}).

The
single-nucleon contributions to the current operator have already been shown to
fulfill the continuity equation in Sec.~\ref{sec:1N:ffparametrization}. 
The leading two-nucleon contributions to the current operator at
order $Q^{-1}$, stemming from the OPE at tree level,
are also well known to fulfill the continuity equation. Further, as already
pointed out in the introduction, the  two-pion exchange
contributions to the current operator at order $Q$ are shown to
be conserved in Ref.~\cite{Kolling:2009iq}. Thus, it remains to verify
the validity of the continuity equation for the order-$Q$
short-range pieces and corrections to the OPE.  

The OPE contributions at the leading one-loop level have been
calculated in~\cite{Kolling:2011mt} {\it without
considering possible contributions induced by  the explicit
time dependence of the imposed UTs in Eqs.~(\ref{betas},\ref{unitS})}.
We have verified that using the choice of unitary
phases $\bar \beta_i$ from Ref.~\cite{Kolling:2011mt} described in
Sec.~\ref{sec:intro}, the OPE contributions to the charge and
current operators up to order $Q^1$ do {\it not} depend on the energy
transfer $k_0$. Thus, the complete expressions for the OPE
contributions to the charge and current operators at order $Q$ are
given in Eqs.~(4.28)-(4.31) of Ref.~\cite{Kolling:2011mt}. 
We now verify the validity of the continuity equation at order $Q^2$ for the
OPE contributions and begin with the longitudinal part of the current operator
\beqa
\vec{k}\cdot\vec{{\fet V}}_{{\rm 2N:}\, 1\pi}^{(Q)}&=&-i\frac{\bar{d}_{18} e \,g_A
  M_\pi^2}{F_\pi^2}[{\fett \tau}^{(1)}\times{\fett
  \tau}^{(2)}]_3\frac{\vec{q}_1\cdot\vec{\sigma}^{(1)}
  \vec{q}_1\cdot\vec{\sigma}^{(2)}}{\vec q_1^2+M_\pi^2}\; +\; (1\leftrightarrow
2)\nn
&=&[\hat{{\fet{H}}}_{\rm 2N: \, 1\pi}^{(Q^2)}, \hat{\fet V}_{{\rm 1N}}^{0\,(Q^{-3})}], 
\eeqa
where the order-$Q^2$ correction to OPE potential is given by
\beqa
{{\fet{H}}}_{\rm 2N: \, 1\pi}^{(Q^2)}&=&\frac{\bar{d}_{18} g_A M_\pi^2}{F_\pi^2} \frac{\vec{q}\cdot \vec{\sigma}^{(1)}
  \vec{q}\cdot \vec{\sigma}^{(2)}}{\vec q^2 + M_\pi^2} {\fett
  \tau}^{(1)}\cdot{\fett \tau}^{(2)},
\eeqa
with $\vec q $ denoting the three-momentum transfer of the nucleons, and the LEC
$\bar{d}_{18}$ parameterizes the deviation from the Goldberger-Treiman relation~\cite{Fettes:1998ud}.
The leading-order single-nucleon charge operator has the form 
\beqa
 {\fet V}_{{\rm 1N}}^{0\,(Q^{-3})}&=&\frac{e}{2}\big(1+\tau_3\big).
 \eeqa
 Here and in what follows, $\vec \sigma^{(i)}$ and $\fett \tau^{(i)}$
 refer to the Pauli spin and isospin matrices of nucleon $i$. 
On the other hand, the longitudinal component of the
energy-transfer-dependent  current is given by
\beqa
\frac{\partial}{\partial k_0}\vec{k}\cdot\vec{{\fet V}}_{{\rm 1N: \, off-shell}}^{(Q)}&=& e \,
{\fet G}_E^{[Q^2]}(Q^2) \,=\,{\fet V}_{{\rm 1N: \, static}}^{0\,(Q^{-1})}\,,\label{OPE:1NstatEqDerk0}
\eeqa
so that
\beqa
\Big[\hat{{\fet{H}}}_{{\rm 2N: \, 1\pi}}^{(Q^0)}, \hat{\fet V}_{{\rm 1N: \, 
    static}}^{0\,(Q^{-1})} - \frac{\partial}{\partial
  k_0}\vec{k}\cdot\hat{\vec{{\fet V}}}_{{\rm 1N: \, off-shell}}^{(Q)}\Big]&=&0.
\eeqa
The double commutator in the continuity
equation~(\ref{Continuityeq:general}) contributes at orders higher than
$Q^2$. Thus, we conclude that the continuity equation for the
contributions of the OPE-range is fulfilled at  order $Q^2$
in the static limit of $m \to \infty$.
Notice that without the $\frac{\partial}{\partial
  k_0}\vec{k}\cdot\hat{\vec{{\fet V}}}_{{\rm 1N: \, off-shell}}^{(Q)}$
contribution, the traditional form of the continuity equation
$\vec{k}\cdot\hat{\vec{{\fet V}}} = [ \hat{{\fet{H}}}, \, \hat{\fet
  V}^0]$ would not be satisfied. 
The validity of the continuity
equation for the leading relativistic corrections at the same chiral
order follows trivially from $\vec{{\fet V}}_{{\rm
    2N:}\, 1\pi , \, 
  1/m}^{(Q)} = 0$ along with the observation
\beq
{\fet V}_{{\rm 1N:} \, 1/m}^{0\,(Q^{-1})} = \frac{\partial}{\partial k_0}
\vec{{\fet V}}_{{\rm 2N:}\, 1\pi}^{(Q^0)} = \vec{{\fet V}}_{{\rm 1N:
 \,    off-shell}}^{(Q^0)} =0\,,
\eeq
which implies that
\beqa
\Big[\hat{{\fet{H}}}_{\rm 2N: \, 1\pi}^{(Q^0)}, \hat{\fet V}_{{\rm 1N:} \,
  1/m}^{0\,(Q^{-1})} \Big] &=&
\bigg[\frac{\hat{\vec{p}}^{\,2}}{2m}, \frac{\partial}{\partial k_0}
\vec{k}\cdot \hat{\vec{{\fet V}}}_{{\rm 2N:}\, 1\pi}^{(Q^0)}\bigg] \nn
&=& \bigg[\hat{{\fet{H}}}_{\rm 2N: \,  1\pi}^{(Q^0)}, \frac{\partial}{\partial
  k_0}\vec{k}\cdot \hat{\vec{{\fet V}}}_{{\rm 1N: \,
    off-shell}}^{(Q^0)}\bigg] \; = \; 0\,.
\eeqa

We now turn to the short-range terms. The leading static contribution to
the current operator is generated by tree-level
diagrams at order $Q^1$ and given in Eq.~(5.3)
of \cite{Kolling:2011mt}. Clearly, the terms in $\vec{k}\cdot
\vec{{\fet V}}_{{\rm 2N: \, 
    cont}}^{(Q)}$ proportional to the LECs $C_i$
coincide with the corresponding contributions in  $[ \hat{{\fet{H}}}_{{\rm 2N:\, 
    cont}}^{(Q^2)}, \, \hat{\fet V}_{{\rm 1N: \, 
    static}}^{0\,(Q^{-3})}]$, while the ones proportional to the
LECs $L_i$ are purely transversal.  Further, 
due to  Eq.~(\ref{OPE:1NstatEqDerk0}), we obtain
\beqa
\Big[\hat{{\fet{H}}}_{{\rm 2N: \, cont}}^{(Q^0)}, \hat{\fet V}_{{\rm 1N: \, 
    static}}^{0\,(Q^{-1})} - \frac{\partial}{\partial
  k_0}\vec{k}\cdot\hat{\vec{{\fet V}}}_{{\rm 1N: \, off-shell}}^{(Q)}\Big]&=&0.
\eeqa
Thus, the continuity equation for the short-range current is fulfilled
at order $Q^2$ in the static limit. Since
\beq
\vec{{\fet V}}_{{\rm 2N: \, cont,} \,  1/m}^{(Q)}  = {\fet V}_{{\rm
    1N:} \, 
  1/m}^{0\,(Q^{-1})} = \frac{\partial}{\partial k_0}
\vec{{\fet V}}_{{\rm 2N: \, cont}}^{(Q^0)}=\vec{{\fet V}}_{{\rm 1N: \, off-shell}}^{(Q^0)}=0\,,
\eeq
the continuity equation  is trivially fulfilled for the short-range relativistic corrections
at order $Q^2$ as well.

\section{Three-nucleon charge operator}
\def\theequation{\arabic{section}.\arabic{equation}}
\label{sec:3N}

\begin{figure}[tb]
\begin{center}  
\includegraphics[width=0.7\textwidth,keepaspectratio,angle=0,clip]{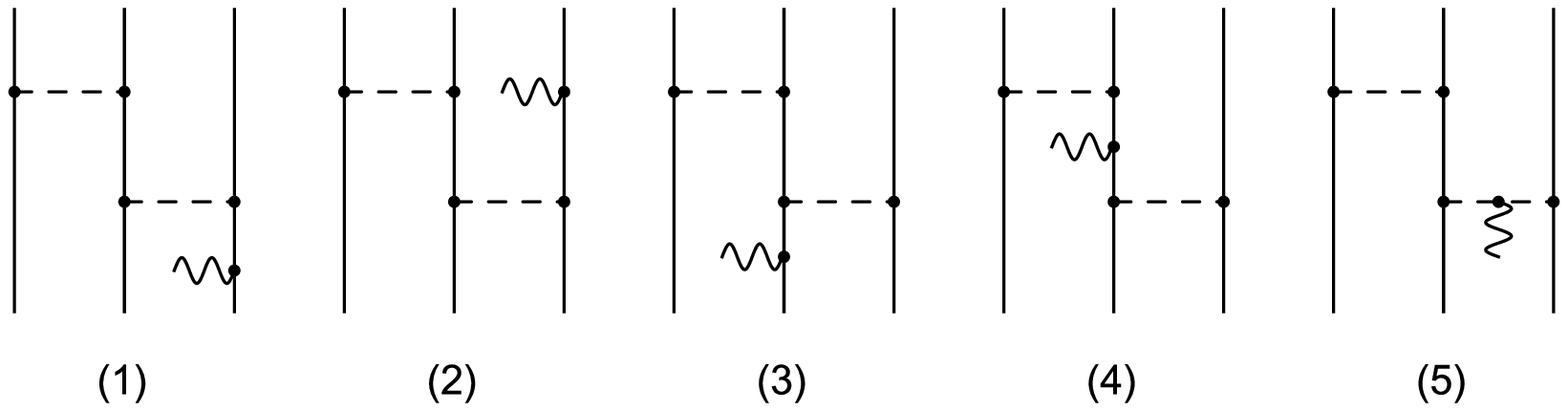}
    \caption{Diagrams generating the $g_A^4$-contribution to the long-range
      three-nucleon electromagnetic charge operator at N$^3$LO. Diagrams resulting from the
      application of the time reversal and permutation operations
      are not shown. Solid, dashed and wiggly lines denote nucleons, pions and photons,
      in order.
\label{fig:3NgA4} 
    }
\end{center}
\vspace{-3mm}
\end{figure}

While the leading contributions to the three-nucleon current appear at
order $Q^2$ which is beyond the accuracy of our study, the
three-nucleon charge operator receives the leading contributions
already at N$^3$LO.
Using the method of UT,  we obtain the
following result for the diagrams $\sim g_A^4$ shown in Fig.~\ref{fig:3NgA4} 
\begin{figure}[tb]
\begin{center} 
\includegraphics[width=0.25\textwidth,keepaspectratio,angle=0,clip]{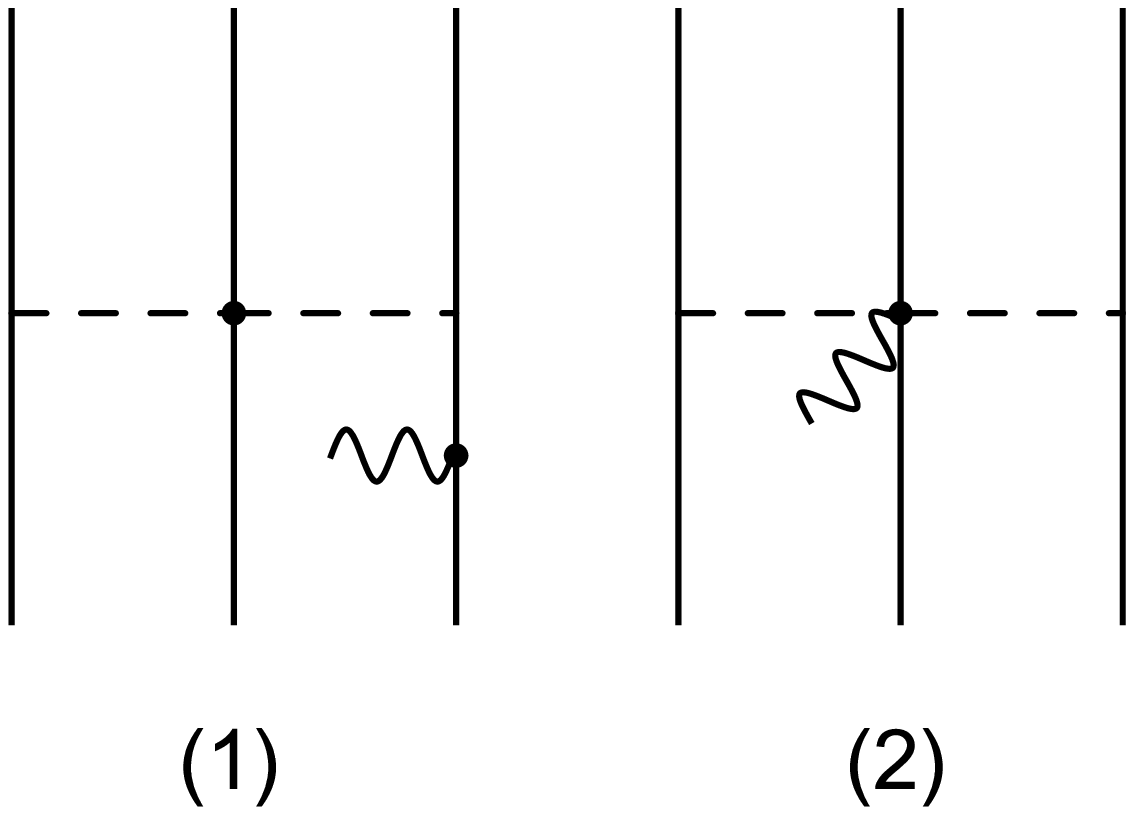}
    \caption{Diagrams generating the $g_A^2$-contribution to the long-range
      three-nucleon electromagnetic charge operator at N$^3$LO. Diagrams resulting from the
      application of the time reversal and permutation operations
      are not shown. For notations, see Fig.~\ref{fig:3NgA4}.
\label{fig:3NgA2} 
    }
\end{center}
\vspace{-3mm}
\end{figure}    

\beqa
\label{3N_1}
{\fet V}_{{\rm 3N: \, \pi} 
}^{0 \, (Q)} &=&-\frac{e\, g_A^4}{8
  F_\pi^4}\frac{\vec{q}_1\cdot\vec{\sigma}^{(1)}}{(\vec
  q_1^2+M_\pi^2)((\vec{q}_1+\vec{q}_2)^2+M_\pi^2)}\bigg(\frac{(\vec{q}_1+\vec{q}_2)\cdot\vec{\sigma}^{(3)}}{(\vec{q}_1+\vec{q}_2)^2+M_\pi^2}+\frac{\vec{q}_3\cdot\vec{\sigma}^{(3)}}{\vec
  q_3^2+M_\pi^2}\bigg)\nn
&\times&\bigg(i[{\fett
  \tau}^{(1)}\times{\fett \tau}^{(3)}]_3 +{\fett \tau}^{(1)}\cdot{\fett \tau}^{(3)}{
  \tau}^{(2)}_3 -{\fett \tau}^{(2)}\cdot{\fett \tau}^{(3)} {
  \tau}^{(1)}_3\bigg)(\vec q_1^2+\vec{q}_1\cdot\vec{q}_2) \nn
&+& 5 \,{\rm permutations},
\eeqa
while the $\sim g_A^2$ contributions visualized in
Fig.~\ref{fig:3NgA2} have the form 
\begin{figure}[tb]
\begin{center}
\includegraphics[width=0.7\textwidth,keepaspectratio,angle=0,clip]{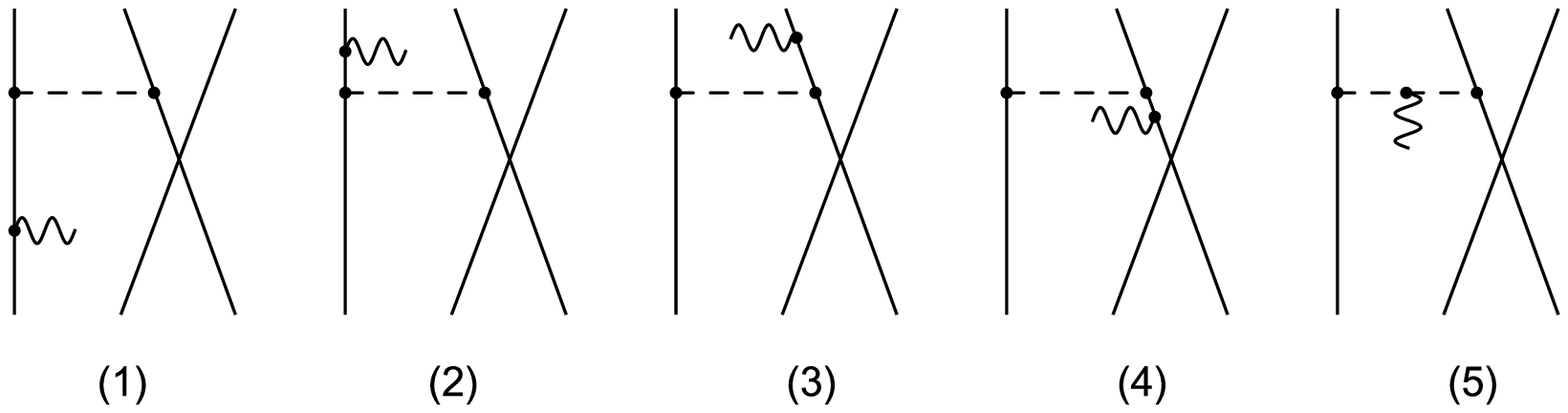}
    \caption{Diagrams generating the $g_A^2$-contribution to the 
      three-nucleon electromagnetic charge operator at N$^3$LO which involve a single
      insertion of the LO NN contact interactions. Diagrams resulting from the
      application of the time reversal and permutation operations
      are not shown.  For notations, see Fig.~\ref{fig:3NgA4}.
\label{fig:3NgA2:contact} 
 }
\end{center}
\vspace{-3mm}
\end{figure}
\beqa
\label{3N_2}
{\fet V}_{{\rm 3N: \, \pi}
}^{0 \, (Q)} &=&\frac{e\,g_A^2}{16
  F_\pi^4}\frac{\vec{q}_1\cdot\vec{\sigma}^{(1)}}{\vec
  q_1^2+M_\pi^2}\bigg(\frac{(\vec{q}_1+\vec{q}_2)\cdot\vec{\sigma}^{(3)}}{\vec{q}_1+\vec{q}_2)^2+M_\pi^2}+\frac{\vec{q}_3\cdot\vec{\sigma}^{(3)}}{\vec
  q_3^2+M_\pi^2}\bigg)\nn
&\times & \big({\fett
  \tau}^{(1)}\cdot{\fett \tau}^{(3)}{\tau}^{(2)}_3-{\fett
  \tau}^{(2)}\cdot{\fett \tau}^{(3)}{\tau}^{(1)}_3\big) \; + \; 5 \,{\rm permutations}.
\eeqa
There are also diagrams involving a single insertion of the
leading-order contact interactions shown in
Fig.~\ref{fig:3NgA2:contact}, which lead to the following result:
\beqa
\label{3N_3}
{\fet V}_{{\rm 3N: \, cont}}^{0\, (Q)} &=&
-[{\fett \tau}^{(1)}\times{\fett\tau}^{(3)}]_3
\frac{e\,g_A^2 C_T}{2 F_\pi^2}
\frac{
(\vec{q}_2+\vec{q}_3)\cdot
    (\vec{\sigma}^{(2)}\times\vec{\sigma}^{(3)})
}{
(\vec{q}_2+\vec{q}_3)^2+M_\pi^2
} \nn
&\times & 
\bigg(\frac{\vec{q}_1\cdot\vec{\sigma}^{(1)}}{\vec q_1^2+M_\pi^2}+\frac{(\vec{q}_2+\vec{q}_3)\cdot\vec{\sigma}^{(1)}}{(\vec{q}_2+\vec{q}_3)^2+M_\pi^2}\bigg)
\; +\;  5 \,{\rm permutations}.
\eeqa
To the best of our knowledge, the three-nucleon contributions to the
charge operator have not been considered before in the framework of
chiral effective field theory. 

\section{Summary and conclusions}
\def\theequation{\arabic{section}.\arabic{equation}}
\label{sec:summary}

\begin{table}[t]
\caption{Chiral expansion of the nuclear electromagnetic current operator up to
  N$^3$LO. LO, NLO, N$^2$LO and N$^3$LO refer to chiral orders $Q^{-3}$,
  $Q^{-1}$, $Q^{0}$ and $Q$, respectively.  The single-nucleon contributions are given in
  Eqs.~(\ref{V1N:Current:nonrel}) and (\ref{singleN:offshell}).
\label{tab_sum_current}}
\smallskip
\begin{tabularx}{\textwidth}{lcrcrcr}
\hline
  \noalign{\smallskip}
 order &&  single-nucleon  &&  two-nucleon  &&
                                                            three-nucleon  
\smallskip
 \\
\hline 
&&&&&& \\[-7pt]
LO  && ---
             && --- && --- \\ [5pt] 
&&&&&& \\[-9pt]
NLO  && $\vec{\fet V}_{{\rm 1N:  \, static}} \;$    \hspace{0.2cm}  &&
                                                             $\vec{\fet
                                                             V}_{{\rm
                                                             2N:  \,
                                                             1\pi}}
                                                             $,
                                                             Eq.~(4.16)
                                          of
                                                             \cite{Kolling:2011mt} && --- \\[2pt]
  && $+ \; \vec{\fet V}_{{\rm 1N:  \, 1/m}} \;$     \hspace{0.36cm}
  &&  && \\
[5pt] 
&&&&&& \\[-9pt]
N$^2$LO && $\vec{\fet V}_{{\rm 1N:  \, static}} \;$
                    \hspace{0.2cm}  && --- && --- \\ [5pt] 
&&&&&& \\[-9pt]
N$^3$LO && $\vec{\fet V}_{{\rm 1N:  \, static}} \;$
                \hspace{0.2cm}   && $\vec{\fet
                                                           V}_{{\rm
                                                           2N:  \,
                                                           1\pi}} $, Eq.~(4.28)
                                          of
                                                             \cite{Kolling:2011mt}
                                            &&  ---
  \\ [2pt]
&&  $+ \; \vec{\fet V}_{{\rm 1N:  \, 1/m}}$ \hspace{0.43cm}  && $+
                                                                  \; \vec{\fet
                                                           V}_{{\rm
                                                           2N:  \,
                                                           2\pi}} $,
                                                              Eq.~(2.18)
                                          of
                                                             \cite{Kolling:2009iq}  
                                            &&  
  \\ [2pt]
&& \hskip -0.35 true cm $+ \; \vec{\fet V}_{{\rm 1N:  \, off-shell}}$&&
  \hskip -0.35 true cm $+
                                                                  \; \vec{\fet
                                                           V}_{{\rm
                                                           2N:  \,
                                                           cont}} $, Eq.~(5.3)
                                          of
                                                             \cite{Kolling:2011mt}&&
                                                             \\ &&&&&&
                                                             \\[-7pt]
                                                             \hline
\end{tabularx}
\end{table}

\begin{table}[t]
\caption{Chiral expansion of the nuclear electromagnetic charge operator up to
  N$^3$LO. LO, NLO, N$^2$LO and N$^3$LO refer to chiral orders $Q^{-3}$,
  $Q^{-1}$, $Q^{0}$ and $Q$, respectively. The single-nucleon contributions are given in
  Eq.~(\ref{V1N:Charge:nonrel}).
\label{tab_sum_charge}}
\smallskip
\begin{tabularx}{\textwidth}{lcrrr}
  \hline
\noalign{\smallskip}
 order &&  single-nucleon  &  two-nucleon  &
                                                            three-nucleon  
\smallskip
 \\
\hline 
&&&& \\[-7pt]
LO && ${\fet V}^0_{{\rm 1N:  \, static}} \;$
             & --- \hspace*{0.5cm}& --- \hspace*{0.5cm}\\ [5pt] 
&&&& \\[-9pt]
NLO &&  ${\fet V}^0_{{\rm 1N:  \, static}} \;$  &--- \hspace*{0.5cm}& --- \hspace*{0.5cm}\\
[5pt] 
&&&& \\[-9pt]
N$^2$LO && ${\fet V}^0_{{\rm 1N:  \, static}} \;$
                   & --- \hspace*{0.5cm}& --- \hspace*{0.5cm}\\ [5pt] 
&&&& \\[-9pt]
N$^3$LO && ${\fet V}^0_{{\rm 1N:  \, static}} \;$
            & ${\fet
                                                           V}^0_{{\rm
                                                           2N:  \,
                                                           1\pi}} $, Eq.~(4.30)
                                          of
                                                             \cite{Kolling:2011mt}\hspace*{0.56cm}
                                            &  ${\fet V}^0_{{\rm 3N:
                                              \, \pi}}$,
                                              Eq.~(\ref{3N_1})  \hspace*{0.20cm}
  \\ [2pt]
&&  $+ \; {\fet V}^0_{{\rm 1N:  \, 1/m}}$ \hspace{0.14cm}  & $+
                                                                  \; {\fet
                                                           V}^0_{{\rm
                                                           2N:  \,
                                                           2\pi}} $,
                                                              Eq.~(2.19)
                                          of
                                                             \cite{Kolling:2009iq}   \hspace*{0.47cm}
                                            &   $+ \; {\fet V}^0_{{\rm 3N:
                                              \, \pi}}$,
                                              Eq.~(\ref{3N_2})  \hspace*{0.20cm}
  \\ [2pt]
&&$+ \; {\fet V}^0_{{\rm 1N:  \, 1/m^2}}$\hspace{0.1cm}  &
 $+
                                                                  \; {\fet
                                                           V}^0_{{\rm
                                                           2N:  \,
                                                           cont}} $, Eq.~(5.6)
                                          of
                                                             \cite{Kolling:2011mt}$^\dagger$
                                            \hspace*{0.31cm}& $+ \; {\fet V}^0_{{\rm 3N:
                                              \, cont}}$,
                                              Eq.~(\ref{3N_3})  
  \\
[2pt]
&&  &
 $+
                                                                  \; {\fet
                                                           V}^0_{{\rm
                                                           2N:  \,
                                                           1\pi , \, 1/m}} $, Eq.~(4.30)
                                          of
                                                             \cite{Kolling:2011mt}&
                                                             \\
                                                             &&&& \\[-7pt]
                                                             \hline
\multicolumn{5}{l}{$^\dagger$ Different conventions are being used in the literature for
  the  leading-order\,  two-nucleon} \\
\multicolumn{5}{l}{contact interactions $\propto C_{S, T}$. To match the
  convention of
  Refs.~\cite{Epelbaum:2014sza,Reinert:2017usi,Epelbaum:2014efa}, the
  factors of $32 F_\pi^2$}   \\
\multicolumn{5}{l}{ in Eq.~(5.7) of \cite{Kolling:2011mt} should be
  replaced by $16 F_\pi^2$.}
\end{tabularx}
\end{table}

In this paper, we have completed the derivation of the nuclear
electromagnetic current
and charge operators to fourth order (N$^3$LO) in the chiral
expansion. The corresponding contributions, obtained using dimensional
regularization, are summarized in Tables~\ref{tab_sum_current} and \ref{tab_sum_charge}. 

The main results of our study can be summarized as follows:
\begin{itemize}
\item
We have analyzed the single-nucleon contributions to the electromagnetic
four-current at the one-loop level using the method of UT and
identified the choice of the unitary
transformations leading to a renormalized result.
We also provide a parametrization of the single-nucleon operators
including the leading and subleading relativistic corrections in
terms of the electromagnetic form factors of the nucleon. Apart from
the well known on-shell pieces, see e.g.~\cite{Marcucci:2015rca},  we found an additional
energy-dependent off-shell contribution to the current operator whose
form is dictated by the renormalizability constraint.  
\item
  We have verified the completeness of the expressions for the
  two-nucleon contributions derived in Refs.~\cite{Kolling:2009iq,Kolling:2011mt} without
  considering possible energy-dependent pieces due to the explicit
  time-dependence of the employed unitary transformations, see
  Ref.~\cite{Krebs:2016rqz} for more details. For the choice of unitary phases
  $\bar \beta_i$ adopted in Refs.~\cite{Kolling:2011mt}, the two-nucleon current
  operator   $\fet V_{\rm 2N}^\mu$ is found to be $k_0$-independent. 
\item
We have worked out the leading three-nucleon contributions to the
charge operator which appears at N$^3$LO.   
\item
With the {\it complete} N$^3$LO result for the electromagnetic current $\fet
V^\mu$ at hand, we were able to explicitly verify the validity of the
modified form of the continuity equation derived in Ref.~\cite{Krebs:2016rqz} at
order $Q^2$ for all one- and two-nucleon contributions.
\end{itemize}
The obtained results can be used to study
electromagnetic properties of nuclei such as the form factors, photo-
and electrodisintegration reactions and the radiative capture
processes with light nuclei in the framework of chiral effective field
theory with fully consistent nuclear forces and currents. This,
however, would require the development and implementation of the
regularization procedure, which is {\it consistent} with the one
employed in the high-precision semilocal chiral nuclear potentials of
Ref.~\cite{Reinert:2017usi}. Work along these lines is in progress.

\vfill
\section*{Acknowledgments}

This work was supported by DFG (SFB/TR 110, ``Symmetries and the Emergence of Structure in
QCD'') and the BMBF  (Grant No. 05P15PCFN1). Further support was provided by the Chinese
Academy of Sciences (CAS) President's International
Fellowship Initiative (PIFI) (grant no. 2018DM0034) and by VolkswagenStiftung
(grant no. 93562). 

\appendix

\section{Unitary transformation to exactly generate the off-shell
  single-nucleon current in Eq.~(\ref{singleN:offshell})}
\label{unitarytr:singenucleon:general}
In this appendix we  construct a unitary transformation,
which
generates the off-shell single-nucleon longitudinal current in Eq.~(\ref{singleN:offshell}).

In order to see how the nuclear operators are affected by time-dependent 
unitary transformations, we perform an additional unitary
transformation of the kind
\beqa
\fet U (t)&=&\exp\big(\fet J(t)\big), \quad \fet J(t)\,=\,\int d^3 x \,v_\mu(\vec{x},
t) \, \fet X^\mu(\vec{x}\,).
\eeqa
where $v_\mu(x)$ is an external vector source and $\fet X_\mu(\vec{x}\,)$ is some
antihermitean operator. The unitary transformation changes the Hamilton
operator in the presence of the vector source to
\beq
\fet W [v] \to \fet U^\dagger(t) \fet W[v] \fet U(t) + \bigg(i\frac{\partial}{\partial
  t}\fet U^\dagger(t)\bigg) \fet U(t)=\fet W[v]+\big[\hat{\fet H}, \,
\hat{\fet J}(t)\big] -  i\frac{\partial}{\partial
  t} \fet J(t) + {\cal O}(v^2),
\eeq
with $\fet H:=\fet W[0]$.
For the vector current operator one obtains\footnote{We adopt here
  the commonly accepted sign convention for the electromagnetic current, which
  differs from the one used in our paper \cite{Krebs:2016rqz}.} 
\beqa
\fet V_\mu(\vec{k})&=&-\frac{\delta \fet W[v]}{\delta \tilde
  v^\mu(\vec{k},k_0)}\Bigg|_{v=0}, \quad \tilde{v}^\mu(\vec{k},k_0) :=\int \frac{d^4
    x}{(2\pi)^4} e^{i k\cdot x} v^\mu(x),
  \eeqa
see \cite{Krebs:2016rqz} for more details,
where the operator $\fet W[v]$ is taken at $t=0$:
\beqa
\fet W[v]=\fet H+\int d^3 x \,v_\mu(\vec{x},0)\, \tilde{\fet
  V}^\mu(\vec{x}\,) \,=\,\fet H +
\int d^4 k\, \tilde{v}_\mu(\vec{k},k_0)\, \fet V^\mu(\vec{k}\,).
\eeqa
Using
\beqa
\fet J(0)&=&\int d^3 x\, v^\mu(\vec{x},0) \, \fet X_\mu(\vec{x}\,)\,=\,\int d^4 k
\,\tilde{v}^\mu(\vec{k},k_0)\, \tilde{\fet X}_\mu(\vec{k}\,), \nn
i\frac{\partial}{\partial t} \fet J(t)\bigg|_{t=0}&=&\int d^3 x\,
i\,\dot{v}^\mu(\vec{x},0) \, \fet X_\mu(\vec{x}\,)\,=\,\int d^4 k\,
k_0 \tilde{v}^\mu(\vec{k},k_0)\, \tilde{\fet X}_\mu(\vec{k}\,),
\eeqa
where 
\beq
\tilde{\fet X}^\mu(\vec{k}\,):=\int d^3 x \,e^{i
  \vec{k}\cdot\vec{x}}\, \fet X^\mu(\vec{x}\,),
\eeq
we obtain
\beqa
\fet V^\mu(\vec{k}\,)\to \fet V^\mu(\vec{k}\,) - \bigg(k_0
\tilde{\fet X}^\mu(\vec{k}\,) - \big[\hat{\fet H}, \, \hat{\tilde{\fet
    X}}^\mu(\vec{k}\,)\big]\bigg).\label{longitudinal:offshell:current}
\eeqa
Thus, choosing
\beqa
\langle p^\prime|\tilde{\fet X}^0(\vec{k}\,)|p\rangle&=&0,\nn
\langle p^\prime|\tilde{\vec{\fet X}}(\vec{k}\,) |p\rangle&=&\chi^{\prime\,\dagger}\bigg(-\vec{k}
\frac{e}{\vec k^2}\bigg[\big({\fet G}_E(\vec k^2)-{\fet
  G}_E(0)\big)+\frac{i}{2m^2}\vec{k}\cdot(\vec{k}_1\times\vec{\sigma})\big({\fet
  G}_M(\vec k^2)\nn
&-&{\fet G}_M(0)\big)\bigg]\bigg)\chi,
\eeqa
we obtain the off-shell term given in
Eq.~(\ref{singleN:offshell}).
It is important to emphasize that the derivation of 
Eq.~(\ref{longitudinal:offshell:current}) does {\it not} rely on 
chiral perturbation theory. No approximations were made to derive this
equation (except for neglecting the contributions involving more than
a single insertion of the external vector source). 

\end{document}